\pdfoutput=1

\documentclass[twocolumn]{aastex631}
\usepackage{amsmath}
\usepackage{graphicx}
\usepackage{mathrsfs}
\usepackage{color}
\usepackage{url}
\usepackage{CJKutf8}
\usepackage{ulem}
\usepackage{wasysym}

\received{2022}
\revised{\today}
\accepted{2022}

\shorttitle{C/2014 UN271}
\shortauthors{Hui et al. 2022}

\begin{document}

\title{
Hubble Space Telescope Detection of the Nucleus of Comet C/2014 UN$_{271}$ (Bernardinelli-Bernstein)
}

\correspondingauthor{Man-To Hui}
\email{mthui@must.edu.mo}

\author{\begin{CJK}{UTF8}{bsmi}Man-To Hui (許文韜)\end{CJK}}
\affiliation{State Key Laboratory of Lunar and Planetary Science, 
Macau University of Science and Technology, 
Avenida Wai Long, Taipa, Macau}

\author{David Jewitt}
\affiliation{Department of Earth, Planetary and Space Sciences, UCLA, 
595 Charles Young Drive East, Los Angeles, CA 90095-1567, USA}

\author{Liang-Liang Yu}
\affiliation{State Key Laboratory of Lunar and Planetary Science, 
Macau University of Science and Technology, 
Avenida Wai Long, Taipa, Macau}

\author{Max J. Mutchler}
\affiliation{Space Telescope Science Institute, Baltimore, 
3700 San Martin Drive, Baltimore, MD 21218, USA}


\newcommand{\UN}{C/2014 UN$_{271}$}

\begin{abstract}

We present a high-resolution observation of distant comet \UN~(Bernardinelli-Bernstein) using the {\it Hubble Space Telescope} on 2022 January 8. The signal of the nucleus was successfully isolated by means of the nucleus extraction technique, with an apparent $V$-band magnitude measured to be $21.64 \pm 0.11$, corresponding to an absolute magnitude of $8.62 \pm 0.11$. The product of the visual geometric albedo with the effective radius squared is $p_V R_n^2$ = 159$\pm$16 km$^2$. If the ALMA observation by \citet{2022arXiv220113188L} refers to a bare nucleus, we  derive a visual geometric albedo of $0.034 \pm 0.008$ and an effective diameter of $137 \pm 15$ km. If dust contamination of the ALMA signal is present at the maximum allowed level (24\%), we find nucleus diameter  $119 \pm 13$ km and albedo of $0.044 \pm 0.011$. In either case, we confirm that \UN~is the largest long-period comet ever detected. Judging from the measured surface brightness profile of the coma, whose logarithmic gradient varies azimuthally between $\sim$1 and 1.7 in consequence of solar radiation pressure, the mass production is consistent with steady-state production but not with impulsive ejection, as would be produced by an outburst. Using aperture photometry we estimated an enormous (albeit uncertain) mass-loss rate of $\sim$10$^3$ kg s$^{-1}$ at a heliocentric distance of $\sim$20 au.

\end{abstract}

\keywords{
comets: individual (\UN) --- methods: data analysis
}

\section{Introduction}
\label{sec_intro}

Long-period comets are conceived to be compositionally some of the most pristine leftovers from the early solar system. For most of their lifetime, they have been stored in the low-temperature environment of the Oort cloud, at the edge of the solar system \citep{1950BAN....11...91O}. Recent years witnessed identifications of several long-period comets active at ultra-large heliocentric distances ($r_{\rm H} \ga 20$ au), implying that the long-period comets may be more thermally processed than previously thought \citep{,2017ApJ...847L..19J,2021AJ....161..188J,2017ApJ...849L...8M,2018AJ....155...25H,2019AJ....157..162H,2021ApJ...921L..37B}. Unlike most comets that are only active within the orbit of Jupiter ($r_{\rm H} \la 5$ au) driven by sublimation of water ice \citep[e.g.,][]{1950ApJ...111..375W}, the cause of activity in distant comets remains unclear. Possible explanations for trans-Jovian activity  include sublimation of supervolatiles such as CO and CO$_2$ \citep[e.g.,][]{2017PASP..129c1001W}, crystallisation of amorphous ice \citep[e.g., 1P/Halley;][]{1992A&A...258L...9P}, and thermal memory from earlier perihelion passage \citep[e.g., Comet Hale-Bopp;][]{2008ApJ...677L.121S}. Before we can use distantly active comets to directly investigate formation conditions of the early solar system, it is of great scientific importance to understand how their activity unfolds at great heliocentric distances.

The recent discovery of \UN~(Bernardinelli-Bernstein) offers us another excellent opportunity to study the distant population of comets. This long-period comet was found in Dark Energy Survey (DES) data at a remarkable inbound heliocentric distance of $r_{\rm H} \approx 29$ au, with additional prediscovery observations from $>$30 au from the Sun and exhibiting an obvious cometary feature at $r_{\rm H} \ga 20$ au \citep{2021ApJ...921L..37B,2021PSJ.....2..236F,2021ATel14733....1K}. According to the orbital solution by JPL Horizons, the current barycentric orbit of \UN~is highly elliptical (eccentricity $e = 0.9993$) with a high perihelion distance of $q = 11.0$ au and a semimajor axis of $a = \left(1.6 \pm 0.2 \right) \times 10^{4}$ au. Amongst many parameters, the size and the albedo of the cometary nucleus are often of the most importance. Recently, \citet{2022arXiv220113188L} reported that the nucleus of the comet, $137 \pm 17$ km in diameter, is the largest amongst all known long-period comets, and has a visual geometric albedo $p_V = 0.049 \pm 0.011$. In this paper, we present our independent study of the nucleus size and albedo of the comet based an observation at a heliocentric distance of $\sim$20 au, detailed in Section \ref{sec_obs}. We present our analysis in Section \ref{sec_rslt} and discussion in Section \ref{sec_dsc}.

\begin{deluxetable*}{cccccccccc}
\tablecaption{Observing Geometry of Comet \UN~(Bernardinelli-Bernstein)
\label{tab_obs}}
\tablewidth{0pt}
\tablehead{
\colhead{Date \& Time (UT)\tablenotemark{a}} & \colhead{Filter} &  \colhead{$t_{\rm exp}$ (s)\tablenotemark{b}} & \colhead{$r_{\rm H}$ (au)\tablenotemark{c}} & \colhead{${\it \Delta}$ (au)\tablenotemark{d}} & \colhead{$\alpha$ (\degr)\tablenotemark{e}} & \colhead{$\varepsilon$ (\degr)\tablenotemark{f}} & \colhead{$\theta_{-\odot}$ (\degr)\tablenotemark{g}} & \colhead{$\theta_{-\bf v}$ (\degr)\tablenotemark{h}} & \colhead{$\psi$ (\degr)\tablenotemark{i}}
}
\startdata
2022 Jan 08 09:24-09:56 & F350LP & 285 & 19.446 & 19.612 & 2.8 & 78.8 & 66.5 & 334.3 & 2.8 \\
\enddata
\tablenotetext{a}{Mid-exposure epoch.}
\tablenotetext{b}{Individual exposure time.}
\tablenotetext{c}{Heliocentric distance.}
\tablenotetext{d}{Comet-{\it HST} distance.}
\tablenotetext{e}{Phase angle (Sun-comet-{\it HST}).}
\tablenotetext{f}{Solar elongation (Sun-{\it HST}-comet).}
\tablenotetext{g}{Position angle of projected antisolar direction.}
\tablenotetext{h}{Position angle of projected negative heliocentric velocity of the comet.}
\tablenotetext{i}{Orbital plane angle (between {\it HST} and orbital plane of the comet).}
\end{deluxetable*}

\begin{figure}[ht!]
\epsscale{1.15}
\plotone{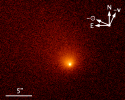}
\caption{{\it HST}/WFC3 F350LP image of comet \UN~(Bernardinelli-Bernstein) median combined from the five individual exposures taken on 2022 January 8. The displayed image is scaled logarithmically and is oriented such that the J2000 equatorial north is up and east is left. Also marked are the directions of the projected antisolar vector ($-\odot$) and the projected negative heliocentric velocity of the comet ($-{\bf v}$). A scale bar of 5\arcsec~in length is shown.
\label{fig:obs}}
\end{figure}

\section{Observation}
\label{sec_obs}




We secured five consecutive images each of 285 s duration in one visit of the comet under General Observer program 16886 using the 2.4 m {\it Hubble Space Telescope} ({\it HST}) and the UVIS channel of the Wide-Field Camera 3 (WFC3) on 2022 January 8. In order to achieve the maximal sensitivity of the facility, we exploited the F350LP filter, which has a peak system throughput of 29\%, an effective wavelength of 585 nm, and a full-width at half maximum (FWHM) of 476 nm. For efficiency we read out only the UVIS2-2K2C-SUB aperture, the $2047 \times 2050$ full quadrant subarray on the UVIS channel with an image scale of 0\farcs04 pixel$^{-1}$ covering a field of view of $81\arcsec \times 81\arcsec$ across. The telescope followed the nonsidereal motion of the comet, resulting in parallax-trailed background sources, despite the great distance of the comet. Image dithering was performed once between the third and fourth exposures so as to mitigate potential impacts from CCD artefacts. The observing geometry of the comet is summarised in Table \ref{tab_obs}.

In the {\it HST} images, the comet possesses a well-defined optocenter inside its bright quasicircular coma of $\sim$4\arcsec~in diameter, with a broad tail of $\ga$15\arcsec~in length directed approximately northeastwards (Figure \ref{fig:obs}).

\section{Analysis}
\label{sec_rslt}

In this section, we present our photometriy to constrain the nucleus of comet \UN~based on our {\it HST} observation. Before carrying out any photometric analysis, we removed cosmic ray hits and hot pixels with the Laplacian cosmic ray rejection algorithm {\tt L.A. Cosmic} by \citet{2001PASP..113.1420V} in {\tt IRAF} \citep{1986SPIE..627..733T}, which successfully rendered us with much cleaner images of the comet while its signal was left untouched. 

\subsection{Direct Photometry}
\label{ssec_phot}

The presence of the bright coma is obviously an obstacle to directly measuring the signal from the nucleus of the comet. However, this enabled us to place upper limits to the contribution of the nucleus. 

The first method we applied was to place a circular aperture of 0\farcs20~(5 pixels) in radius at the centroid of the comet in each of the five individual exposures, regard the measured signal as being all from the nucleus, and determine the sky background using a concentric annulus having inner and outer radii of 8\arcsec~and 40\arcsec, respectively, where contamination from the dust environment of the comet is completely negligible. We thereby obtained the apparent $V$-band magnitude of the region enclosed by the 0\farcs20~radius aperture to be $m_V = 21.10 \pm 0.03$, in which the reported uncertainty is the standard deviation on the repeated measurements. Since the measured signal has contributions from both the nucleus and the surrounding coma enclosed by the aperture, the apparent magnitude of the nucleus must be fainter than the measured one. To correct for the observing geometry, we simply assumed a linear phase function with a slope of $\beta_{\alpha} = 0.04 \pm 0.02$ mag deg$^{-1}$ appropriate for comets at small phase angles \citep[e.g.,][]{2004come.book..223L}. The result is highly unlikely to be altered greatly by the actual phase function, which is observationally unconstrained, in that the phase angle of the comet during our {\it HST} observation was merely 2\fdg8. Accordingly we estimate an uncertainty of $\sim\pm$0.06 introduced by the phase function. 

We computed the absolute magnitude of the nucleus from
\begin{equation}
H_{{\rm n}, V} = m_{{\rm n}, V} - 5 \log \left(r_{\rm H} {\it \Delta}\right) - \beta_{\alpha} \alpha
\label{eq_mabs},
\end{equation}
\noindent where the subscript ``n" denotes parameters for the nucleus, $r_{\rm H}$ and $\it \Delta$ are respectively the heliocentric and cometocentric distances expressed in au, and $\alpha$ is the phase angle in degree. Substituting, we found that the nucleus of the comet must have $H_{{\rm n}, V} > 8.08 \pm 0.03$, in which the uncertainty is the standard error. The geometric albedo and the radius of the nucleus are directly related to the absolute magnitude by
\begin{equation}
p_{V} R_{\rm n}^2 = 10^{0.4 \left(m_{\odot,V} - H_{{\rm n}, V} \right)} r_{\oplus}^{2}
\label{eq_pR2},
\end{equation}
\noindent where $p_V$ is the geometric albedo in the $V$ band, $R_{\rm n}$ is the nucleus radius, and $m_{\odot, V} = -26.76 \pm 0.03$ is the apparent $V$-band magnitude of the Sun at heliocentric distance $r_{\oplus} = 1$ au \citep{2018ApJS..236...47W}. Inserting numbers, we found $p_{V} R_{\rm n}^2 \la \left(2.6 \pm 0.1\right) \times 10^2$ km$^2$. Unfortunately the nucleus size of the comet is subjected to the assumption of the geometric albedo. However,  \citet{2022arXiv220113188L} lately reported $p_V = 0.049 \pm 0.011$ and $R_{\rm n} = 69 \pm 9$ km using their ALMA observation in combination with the optical measurements by \citet{2021ApJ...921L..37B}. If their assumption that the coma contamination was negligible in the ALMA data is valid (but see Section \ref{ssec_nucsz}), the reported size of the nucleus will be trustworthy, because the thermal emission measures $R_{\rm n}^{2}$ and almost has no dependency upon the albedo. Therefore, assuming the aspect angles are not too different between the ALMA and {\it HST} observations, we found an upper limit to the geometric albedo of the nucleus to be $p_V < 0.055 \pm 0.014$, contingent on the nucleus size derived by \citet{2022arXiv220113188L}.

As the coma is apparently bright in the {\it HST} observation, the first method only provides a coarse upper limit to the albedo of comet \UN. Thus, we adopted the second method to better constrain the parameter, in which the contribution from the coma was partially corrected. We still applied the same circular aperture of 0\farcs20 in radius at the centroid of the comet. However, the coma in the contiguous annular region up to 0\farcs28 from the centroid was measured and treated as the background value to be subtracted from the central aperture. The resulting flux measured by the aperture is still an upper limit to the counterpart from the nucleus. This is because this method underestimates the surface brightness of the coma in the central aperture, but it nevertheless provides a better constraint than does the first method, in which no correction was attempted whatsoever. We found the resulting apparent $V$-band magnitude to be $m_V = 21.22 \pm 0.03$, corresponding to $H_{{\rm n},V} > 8.20 \pm 0.03$, and $p_V < 0.050 \pm 0.012$, if still assuming the nucleus size reported by \citet{2022arXiv220113188L}. In comparison, \citet{2022arXiv220113188L} reported the exact geometric albedo of the comet, rather than an upper limit, to be $p_V = 0.049 \pm 0.011$, which is indistinguishable from what we obtained from the second method. We refrain from the relevant discussion until in Section \ref{sec_dsc}.

\begin{figure*}
\epsscale{1.1}
\plotone{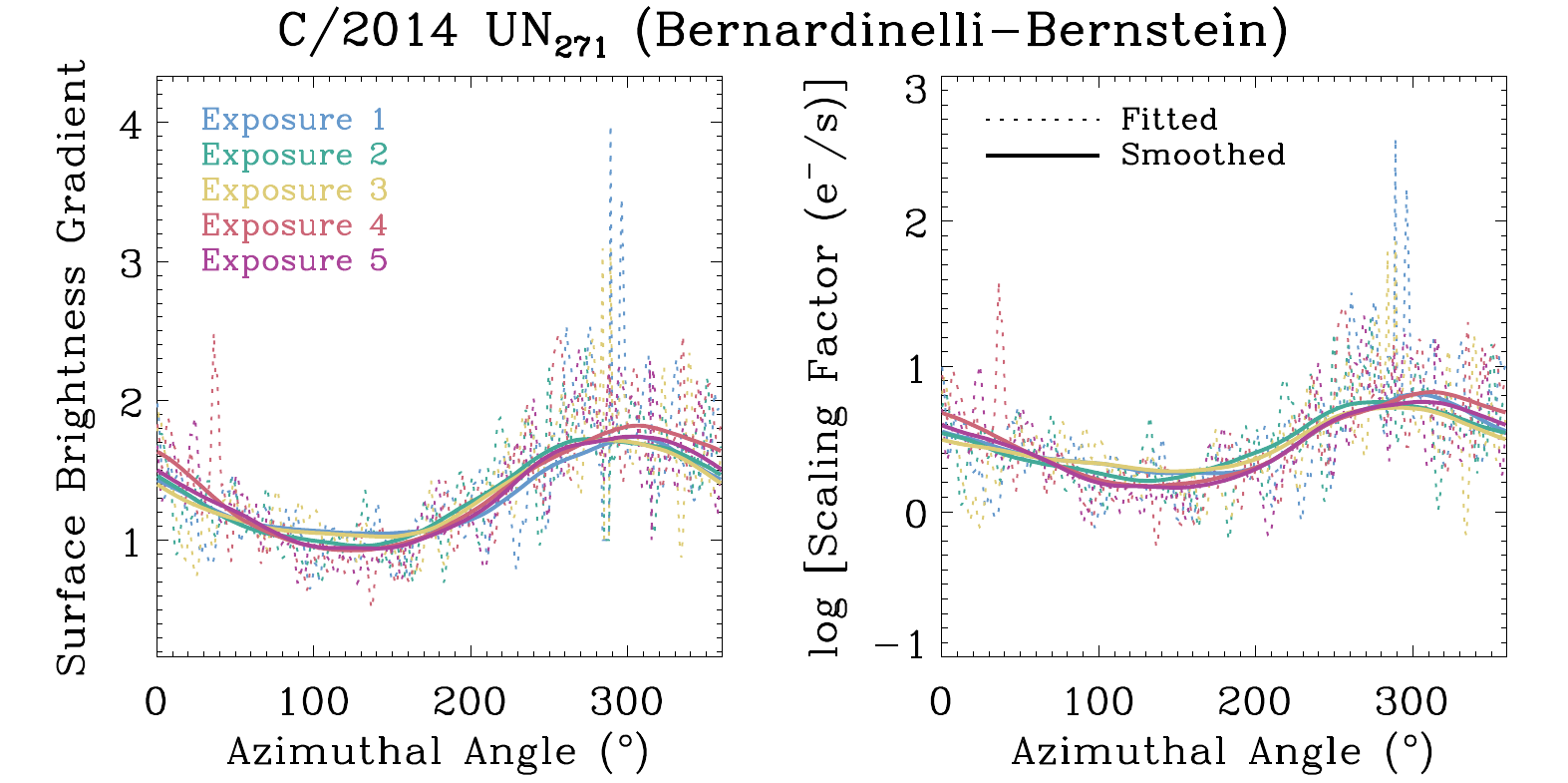}
\caption{Best-fitted (dotted lines) and smoothed (solid lines) logarithmic surface brightness gradient and the scaling factor of the coma both as functions of the azimuthal angle. Results from different individual exposures are distinguished by colors, as indicated in the legend in the left panel. The surface brightness profile of the comet in annular regions between 0\farcs24 and 0\farcs80 from the optocenter in the individual exposures was used for the best fits.
\label{fig:bpars}}
\end{figure*}

\begin{figure*}
\epsscale{1.1}
\plotone{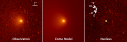}
\caption{Brief illustration of how the nucleus extraction technique was applied for the second {\it HST}/WFC3 exposure as an example. The coma model (middle panel) was obtained by means of fitting the surface brightness profile of the observed image (left panel), followed by subtracting the former from the latter, unveiling a stellar source at the original centroid of the comet in the residual image (right panel), which we interpreted as the nucleus of comet \UN. A 1\arcsec~scale bar and the cardinal directions, along with the directions of the projected antisolar vector and the projected negative heliocentric velocity of the comet are marked.
\label{fig:im_comp}}
\end{figure*}

\begin{figure*}
\plotone{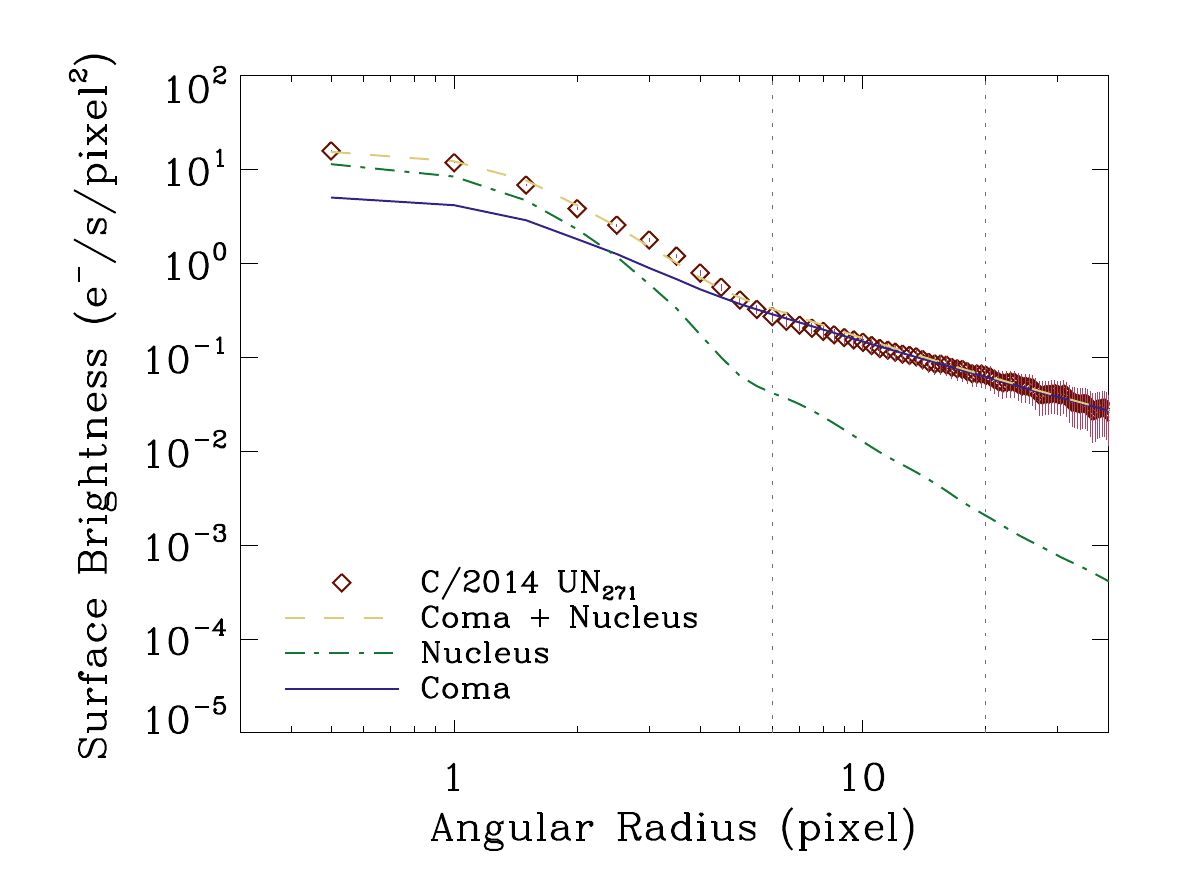}
\caption{Radial profile comparison between the coma (violet solid line), the nucleus (green dashed-dotted line), and the total (yellow dashed line) models, and the observation (red diamonds with error bars given) plotted on a log-log scale for the second individual {\it HST} exposure as an example. Results from the other four exposures are visually similar and are therefore not displayed separately for brevity. The two grey vertical dotted lines mark the annular radius range (6-20 pixels, or 0\farcs24-0\farcs80) within which the surface brightness profile of the coma was fitted.
\label{fig:rprof}}
\end{figure*}

\subsection{Nucleus Extraction}
\label{ssec_nucext}

Given the ultrastable point-spread function (PSF) and the supreme spatial resolution and sensitivity of the {\it HST}/WFC3 camera, we opted to employ the nucleus extraction technique, which has been successfully applied for a number of comets previously observed by {\it HST} \citep[e.g.,][]{1998A&A...335L..25L,1998A&A...337..945L,2009A&A...508.1045L,2011MNRAS.412.1573L} and systematically evaluated \citep{2018PASP..130j4501H}. The basic idea of the technique is to remove the contamination of the coma by means of fitting its surface brightness profile and extrapolating inwards to the near-nucleus region, assuming that the coma is optically thin, such that the signal from the coma and that from the nucleus are separable. The surface brightness of the coma was fitted by an azimuthally dependent power-law model. We expressed the surface brightness of the comet as a function of the angular distance to the nucleus ($\rho$) and the azimuthal angle ($\theta$) in the sky plane as
\begin{align}
\nonumber
{\it \Sigma}_{\rm m} \left(\rho, \theta\right) & = \left[k_{\rm n} \delta\left(\rho\right) + k_{\rm c} \left(\theta\right)\left(\frac{\rho}{\rho_0}\right)^{-\gamma\left(\theta\right)}\right] \ast \mathcal{P}\\
& = k_{\rm n} \mathcal{P} + \left[k_{\rm c} \left(\theta\right) \left(\frac{\rho}{\rho_0}\right)^{-\gamma\left(\theta\right)}\right] \ast \mathcal{P}
\label{eq_Sm}.
\end{align}
\noindent Here, $k_{\rm n}$ and $k_{\rm c}$ are the scaling factors for the nucleus and coma, respectively, $\delta$ is the Dirac delta function, $\gamma$ is the logarithmic surface brightness gradient of the coma, $\mathcal{P}$ is the normalised PSF kernel, $\rho_0 = 1$ pixel is a normalisation factor to guarantee that the two scaling factors share the same unit, and the symbol $\ast$ is the convolution operator. 

We followed the procedures detailed in \citet{2018PASP..130j4501H} to extract the nucleus signal from our {\it HST} data. Basically, we fitted the surface brightness profile of the coma in azimuthal segments of 1\degr~over some annular region where the contribution from the nucleus is negligible in the individual exposures. Smoothing of the best-fit parameters for the coma was carried out so as to alleviate fluctuations due to uncleaned artefacts caused by cosmic ray hits (Figure \ref{fig:bpars}), followed by extrapolating the surface brightness profile inwards to the near-nucleus region and convolution with the {\it HST}/WFC3 PSF model generated by {\tt TinyTim} \citep{2011SPIE.8127E..0JK}. Subtraction of the coma model from the observed image revealed a well-defined stellar source around the original centroid of the comet in the residual image, which was measured to have a FWHM of $0\farcs071 \pm 0\farcs004$ (or $1.8 \pm 0.1$ pixels), in line with the FWHM of the PSF model by {\tt TinyTim}, therefore interpreted as the nucleus of the comet (Figure \ref{fig:im_comp}). We then fitted the PSF model to the source, whereby we obtained the scaling factor $k_{\rm n}$ as the total flux for the nucleus using aperture photometry. Comparisons between radial brightness profiles of the observation and the models are plotted in Figure {\ref{fig:rprof}}.

To test the reliability of the results, we varied a number of parameters, including the subsampling factor, the fitted region, and the smoothed angle bins, only to find that the variation is always only $\la$10\% of the measured flux, no greater than the standard deviation of the repeated measurements. Therefore, we used the latter as the uncertainty of the nucleus flux obtained from the nucleus extraction technique, although this most likely overestimates the actual error. 

It is known that the nucleus extraction technique produces systematic biases in determination of nucleus signal that are difficult to correct, and that it can even fail on a few occasions \citep{2018PASP..130j4501H}. In order to ascertain how our result might be biased by the technique, we assessed the ratio between the nucleus flux and the total flux measured with a 0\farcs60 radius circular aperture, which was found to be always $\ga$30\%, falling into a regime where the bias is totally negligible \citep{2018PASP..130j4501H}. Therefore, we are confident that the signal of the nucleus determined from our {\it HST} observation on comet \UN~is robust.

The result is that we found the apparent $V$-band magnitude of the nucleus to be $m_{{\rm n}, V} = 21.64 \pm 0.11$. Substitution into Equation (\ref{eq_mabs}) yields $H_{{\rm n}, V} = 8.62 \pm 0.11$, which is clearly fainter than what \citet{2021ApJ...921L..37B} reported based on their optical observations \citep[$H_{{\rm n},V} = 8.21 \pm 0.05$, converted from the Sloan bands;][]{ 2022arXiv220113188L}, and corresponds to $p_{V} R_{\rm n}^2 = \left(1.59 \pm 0.16\right) \times 10^{2}$ km$^2$ yielded by Equation (\ref{eq_pR2}). Still adopting the nucleus size reported by \citet{2022arXiv220113188L}, we determine a nucleus geometric albedo of $p_V = 0.034 \pm 0.009$, in which the uncertainty was properly propagated from all measured and reported errors. Our result suggests a lower albedo for the nucleus surface, because it is possible that the photometry by \citet{2021ApJ...921L..37B} is contaminated by the dust environment of comet \UN. Nonetheless, the albedo we derived is unremarkable in comparison to those of other cometary nuclei \citep[distributed in a narrow range of $p_V \approx 0.02$-0.06;][]{2004come.book..223L}.

\begin{figure}
\epsscale{1.2}
\plotone{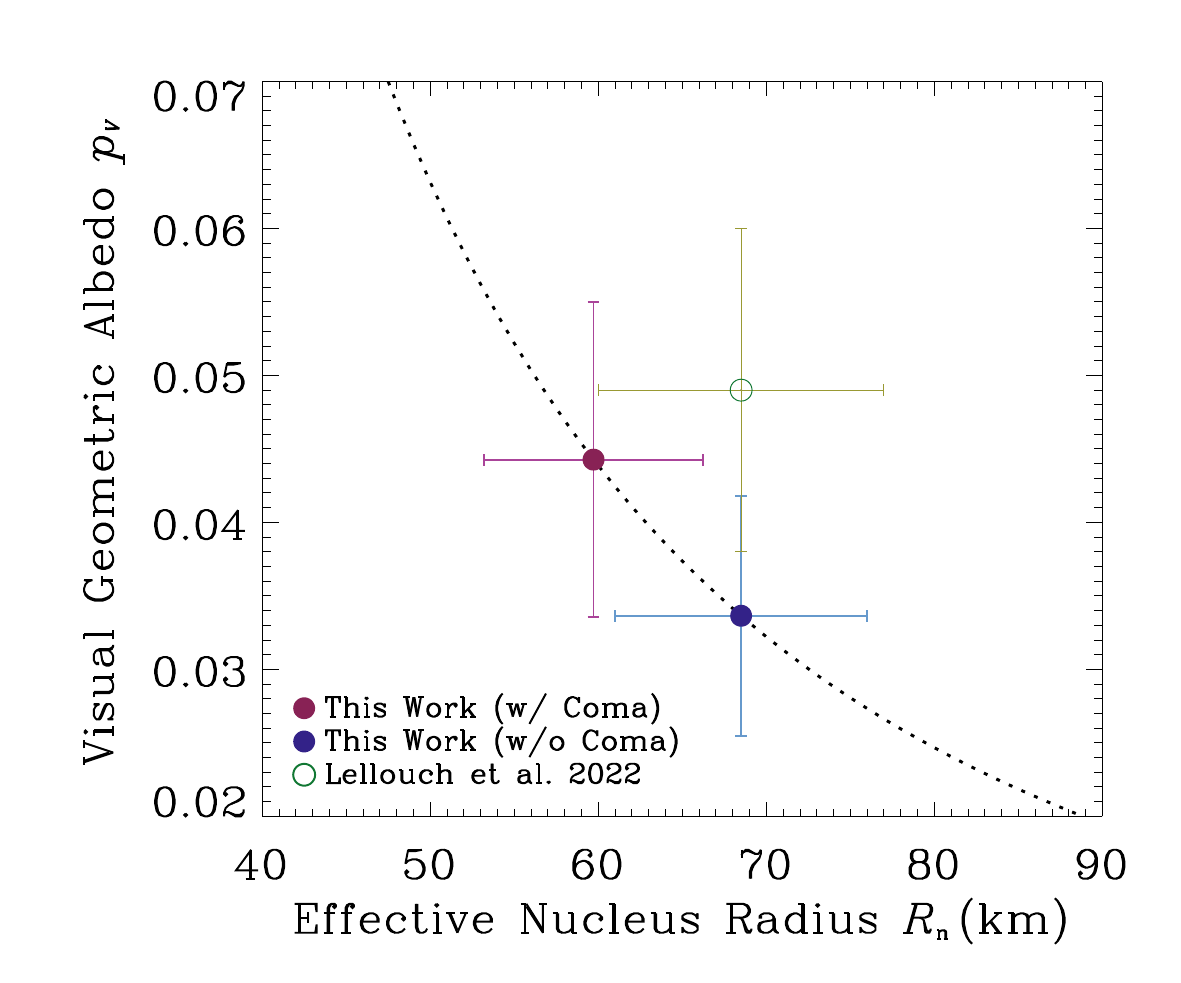}
\caption{Geometric albedo vs.~effective  radius of the nucleus of C/2014 UN$_{271}$.  The dotted line shows $p_V R_{\rm n}^2 =159$ km$^2$, as found   from our HST photometry.  The nucleus parameters must fall on this line.  The hollow green circle marks the measurement by \citet{2022arXiv220113188L}, which relies upon large-aperture  (more likely coma-contaminated) photometry of the nucleus by \citet{2021ApJ...921L..37B}. The filled blue and purple circles show two solutions using the new $p_VR_n^2$ constraint both with and without the $\sim$24\% dust contamination of the 230 GHz thermal signal allowed by \citet{2022arXiv220113188L}. 
\label{fig:comp_pR2}}
\end{figure}

\section{Discussions}
\label{sec_dsc}

\subsection{Nucleus Size}
\label{ssec_nucsz}

Our analysis of the {\it HST} observation of comet \UN~provided us with an estimate of its nucleus absolute magnitude $\sim$0.41 mag fainter than the result by \citet{2021ApJ...921L..37B}, presumably as a result of coma contamination in the large aperture optical photometry used by these authors.  If there is no dust contamination of the 233 GHz ALMA signal, the nucleus albedo must be  lower  than the one derived by \citet{2022arXiv220113188L}, as shown by the hollow green and filled blue circles in Figure \ref{fig:comp_pR2}. 

However, \citet{2022arXiv220113188L} concluded that up to $\sim$24\% of the 233 GHz continuum flux could be from an unseen dust coma in their data. In this  case, we estimate that the effective diameter of the comet would be reduced to $119 \pm 15$ km and the geometric albedo increased to  $p_V = 0.044 \pm 0.012$, shown as a filled purple circle in Figure \ref{fig:comp_pR2}.  In order to affect the 233 GHz cross-section, dust in the comet would need to be large. Two observations suggest that the coma might indeed be rich in large grains. Firstly, independent observations of other long-period comets (notably C/2017 K2; \citep[e.g.,][]{2017ApJ...847L..19J,2018AJ....155...25H,2019AJ....157...65J} have convincingly revealed that submillimeter and larger dust grains are produced at great heliocentric distances.  Secondly, based on a  syndyne-synchrone computation, \citet{2021PSJ.....2..236F} deduced that C/2014 UN$_{271}$ has been ejecting submillimeter sized and larger dust grains  for years prior to the epoch of observation.  The driver of mass loss at distances $\sim$20 au (and potentially at much larger distances; \cite{2022ApJ...924...37B}) is presumably the sublimation of carbon monoxide.

An additional factor which might affect estimates of the albedo is the rotation of the nucleus, resulting in cross-sections different between the ALMA and {\it HST} observations. However, most known cometary nuclei have aspect ratios $\la$2:1 \citep[e.g.,][]{2004come.book..223L} and so we do not expect rotational effects in nucleus photometry larger than a factor of two.

In short, our improved estimate of the absolute magnitude shows that the nucleus of \UN~is slightly smaller or slightly darker than found by  \citet{2022arXiv220113188L} but we strongly confirm their result that the nucleus is larger than any other previously measured long-period cometary nucleus. Figure \ref{fig:comp_pR2} is shown as a comparison of the results.

\begin{figure}
\epsscale{1.2}
\plotone{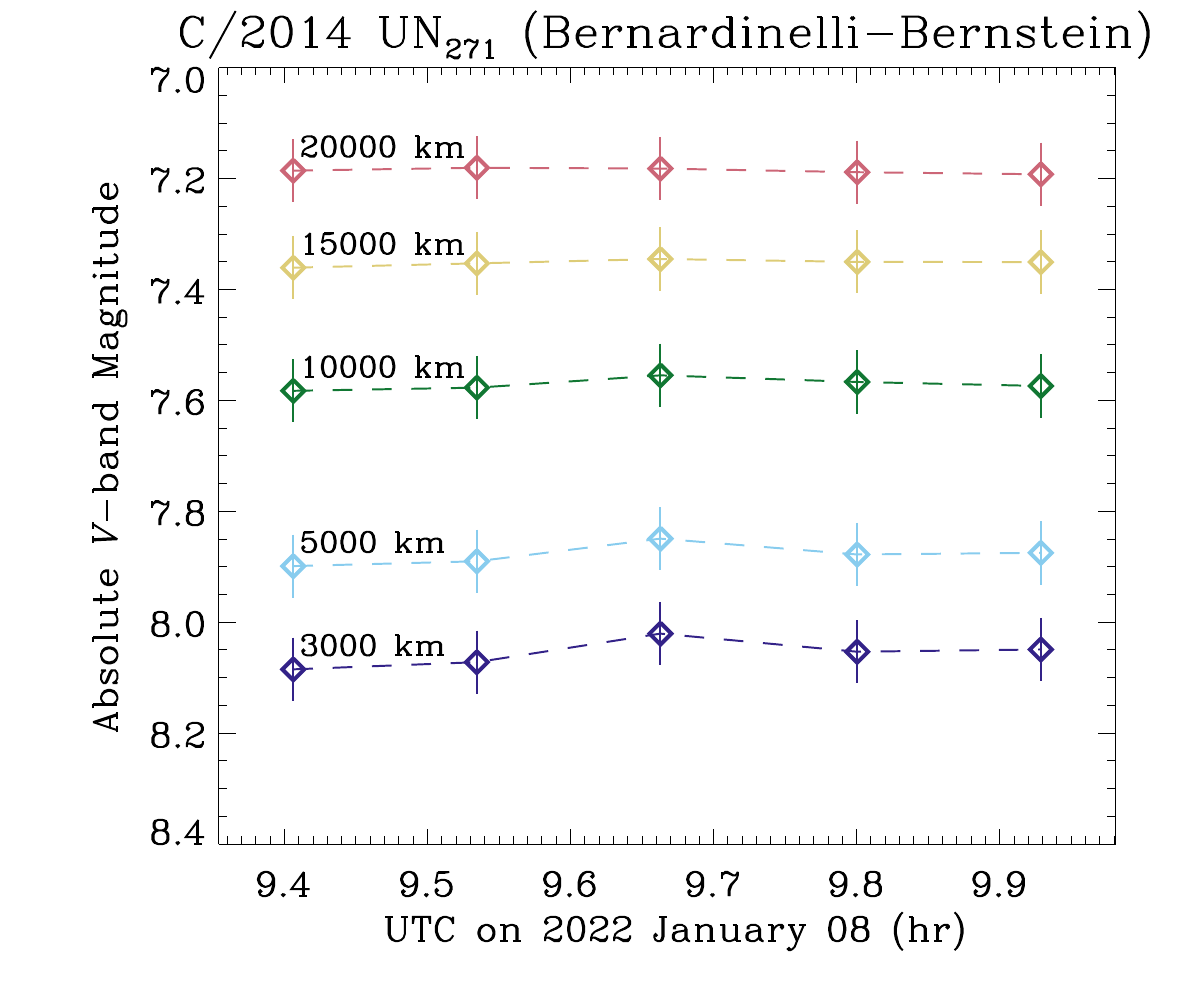}
\caption{Absolute $V$-band magnitude as a function of time in UTC on 2022 January 8 for each of the fixed linear apertures, distinguished by colors. The radii of the apertures are explicitly labelled on the plot. The reported errors are dominated by the uncertainty in the assumed phase function. 
\label{fig:dust}}
\end{figure}

\subsection{Mass Loss}
\label{ssec_ml}

We measured the logarithmic surface brightness gradient of the coma in Section \ref{ssec_nucext}, which allows for qualitative diagnosis of the observed activity. In steady state, the surface brightness gradient of the coma is expected to be $\gamma= 1$ but will be steepened to $\sim$1.5 by the solar radiation pressure \citep{1987ApJ...317..992J}. Indeed, we found that the surface brightness gradient of the coma lies within a range between $\sim$1 and 1.7 (see Figure \ref{fig:bpars}), consistent with the coma produced in steady state. The result that the gradient is generally steeper around the azimuthal angles facing towards the Sun and is shallower otherwise is expected.

In addition to characterising the properties of the nucleus of comet \UN, we also performed photometry of the comet in multiple fixed linear circular apertures aiming to measure its coma. The background was determined in the same fashion as in the first method in Section \ref{ssec_phot}. For correction of the observing geometry, we still assumed a linear phase function with slope $\beta_{\alpha} = 0.04 \pm 0.02$ mag deg$^{-1}$, which is also appropriate for cometary dust at small phase angles \citep[e.g.,][and citations therein]{2004come.book..577K}. We plot the measurements in Figure \ref{fig:dust}, in which the errors are primarily attributed to the uncertainty in the assumed phase function.

In the following we estimate the mass loss of the comet using the largest fixed linear aperture in the order-of-magnitude manner. Presuming that the total cross-section of the dust particles having mean radius $\bar{a}_{\rm d} \sim 0.1$ mm was ejected in steady state at speeds $v_{\rm ej} \sim 10$ m s$^{-1}$ \citep{2021PSJ.....2..236F}, we can then relate the mass-loss rate to the measured absolute magnitude of the dust by
\begin{equation}
\overline{\dot{M}}_{\rm d} \sim \frac{\pi \rho_{\rm d} \bar{a}_{\rm d} v_{\rm ej} r_{\oplus}^2}{\ell p_{V}} 10^{0.4 \left(m_{\odot, V} - H_{{\rm d}, V}\right)}
\label{eq_ml},
\end{equation}
\noindent in which the subscript ``d" denotes parameters of the dust grains, $\rho_{\rm d} \sim 1$ g cm$^{-3}$ is the nominal bulk density of the dust grains, and $\ell = 2 \times 10^{4}$ km is the projected radius of the largest aperture we used to measure the coma. By substitution, we find the dust mass-loss rate $\overline{\dot{M}}_{\rm d} \sim 10^{3}$ kg s$^{-1}$. In comparison, distant comet C/2017 K2 (PANSTARRS) was estimated to exhibit a dust mass-loss rate of $\sim$10$^2$ kg s$^{-1}$ at $r_{\rm H} \la 20$ au \citep{2017ApJ...847L..19J,2018AJ....155...25H,2021AJ....161..188J}, while \citet{2008ApJ...677L.121S} reported $\sim$10$^3$ kg s$^{-1}$ for comet Hale-Bopp at similar heliocentric distances on the outbound leg of its orbit. Yet none of the aforementioned values are better than order-of-magnitude estimates.

\section{Summary}
\label{sec_sum}

We employed the {\it Hubble Space Telescope} to observe the distant active comet \UN~(Bernardinelli-Bernstein) on 2022 January 8. The key conclusions are:

\begin{enumerate}
\item The apparent $V$-band magnitude of the bare cometary nucleus was measured using  high resolution HST data and a profile fitting technique to be $21.64 \pm 0.11$, corresponding to an absolute magnitude of $8.62 \pm 0.11$. The nucleus must satisfy $p_V R_{\rm n}^2 = \left(1.59 \pm 0.16 \right) \times 10^{2}$ km$^2$, where $p_V$ and $R_n$ are the geometric albedo and effective nucleus radius, respectively. 

\item Assuming that the ALMA photometry by \citet{2022arXiv220113188L} is free from any contamination from the dust coma, we estimated the visual geometric albedo and the effective radius of the nucleus to be $p_V = 0.034 \pm 0.008$ and $R_{\rm n} = 69 \pm 8$ km, respectively. If the maximal $\sim$24\% contamination of the ALMA flux contributed by  coma is considered, we instead derive $p_V = 0.044 \pm 0.011$ and $R_{\rm n} = 60 \pm 7$ km.  The nucleus likely lies between these extremes.

\item The logarithmic surface brightness gradient of the coma varies between $\gamma \sim 1$ and 1.7 depending on the azimuthal angle, indicating that the dust grains are ejected in a protracted rather than an impulsive manner.

\item From the photometric measurements of the coma, we estimated the dust mass-loss rate of the comet to be $\sim$10$^{3}$ kg s$^{-1}$ at heliocentric distance $r_{\rm H} \sim 20$ au.

\end{enumerate}

\begin{acknowledgements}
This research is based on observations from program GO 16886 made with the NASA/ESA {\it Hubble Space Telescope} obtained from the Space Telescope Science Institute, which is operated by the Association of Universities for Research in Astronomy, Inc., under NASA contract NAS 5–26555. MTH appreciates great support and encouragement from Kiwi.
\end{acknowledgements}

\vspace{5mm}
\facilities{{\it HST}}

\software{{\tt IDL}, {\tt IRAF} \citep{1986SPIE..627..733T}, {\tt L.A. Cosmic} \citep{2001PASP..113.1420V}, {\tt TinyTim} \citep{2011SPIE.8127E..0JK}}

\end{document}